\documentclass[runningheads]{llncs}
 
 \usepackage{url}
 \usepackage{hyperref}
 \usepackage{eulervm}
 \usepackage{microtype}
 \usepackage{graphicx} \usepackage{amsmath}
\title{Making Math Searchable in Wikipedia}
\author{Moritz Schubotz}
\newcommand{\LaTeXML}{{\sc LaTeXML}}
\newcommand{\MathML}{{\sc MathML}}

\urldef{\mailsa}\path|schubotz@tu-berlin.de|

\institute{Technische Universit{\"a}t Berlin, School of EECS (IV), DIMA-Group\\
Einsteinufer 17, 10587 Berlin, Germany\\
\mailsa\\
\url{http://www.formulaesearchengine.com}}

\begin{document}
\maketitle
\begin{abstract}
Wikipedia, the world largest encyclopedia contains a lot of knowledge that is expressed as formulae exclusively.
Unfortunately, this knowledge is currently not fully accessible by intelligent information retrieval systems.
This immense body of knowledge is hidden form value-added services, such as search.
In this paper, we present our MathSearch implementation for Wikipedia that enables users to perform a combined text
and fully unlock the potential benefits. 
\end{abstract}
\section{Introduction}
Over the past decades almost all published knowledge has been converted into a digital form and is for the most part available free of charge  in the global data network.
Major achievements in information retrieval have contributed to the fact that a majority of humans in the USA use conventional search engines on a daily basis~\cite{Purcell2012}. 
For example, the English Wikipedia project that includes of 4 million articles and has 10.5 over million visits per hour~\cite{ErikZachte}, contains a large portion the existing knowledge.
Especially in mathematics, science and engineering disciplines, crucial parts of knowledge are conveyed in formulae only. 
Unfortunately, these formulae are not taken into account by current (text-based) search engines. There are two main reasons.
Firstly, the formulae are inaccessible, since a large portion of them are, as in the case of Wikipedia, stored as pictures only. 
Secondly, previous efforts in math-oriented search engines based on string search (i.e. formulas expressed using alphanumeric string) have had little success.

However, there is a wide range of applications for mathematical search engines. On the one hand, a well designed math aware search engine for scientific publications helps to avoid that the same things get reinvent over and over. Furthermore it may help junior researchers that are not aware of all the keywords, to get a faster overview about a new topic that they explore. Furthermore, it helps to network communities orthogonally to their field of applications, if they use different technical terms but, try to solve similar problems from a mathematical point of view.
But even beyond the academic activities, application like enterprise search, patent application and technical consulting services will benefit and increase profit by the application of math aware search engines.

In this paper, we demonstrate the power of a combined text and formula search suitable for Wikipedia.
Next, we treat the conversion and indexing of formula section and then present our  combined text and formula search interface in the subsequent section.

\section{Math Parsing in Mediawiki}
\begin{figure}[ht]
\includegraphics[width=\textwidth]{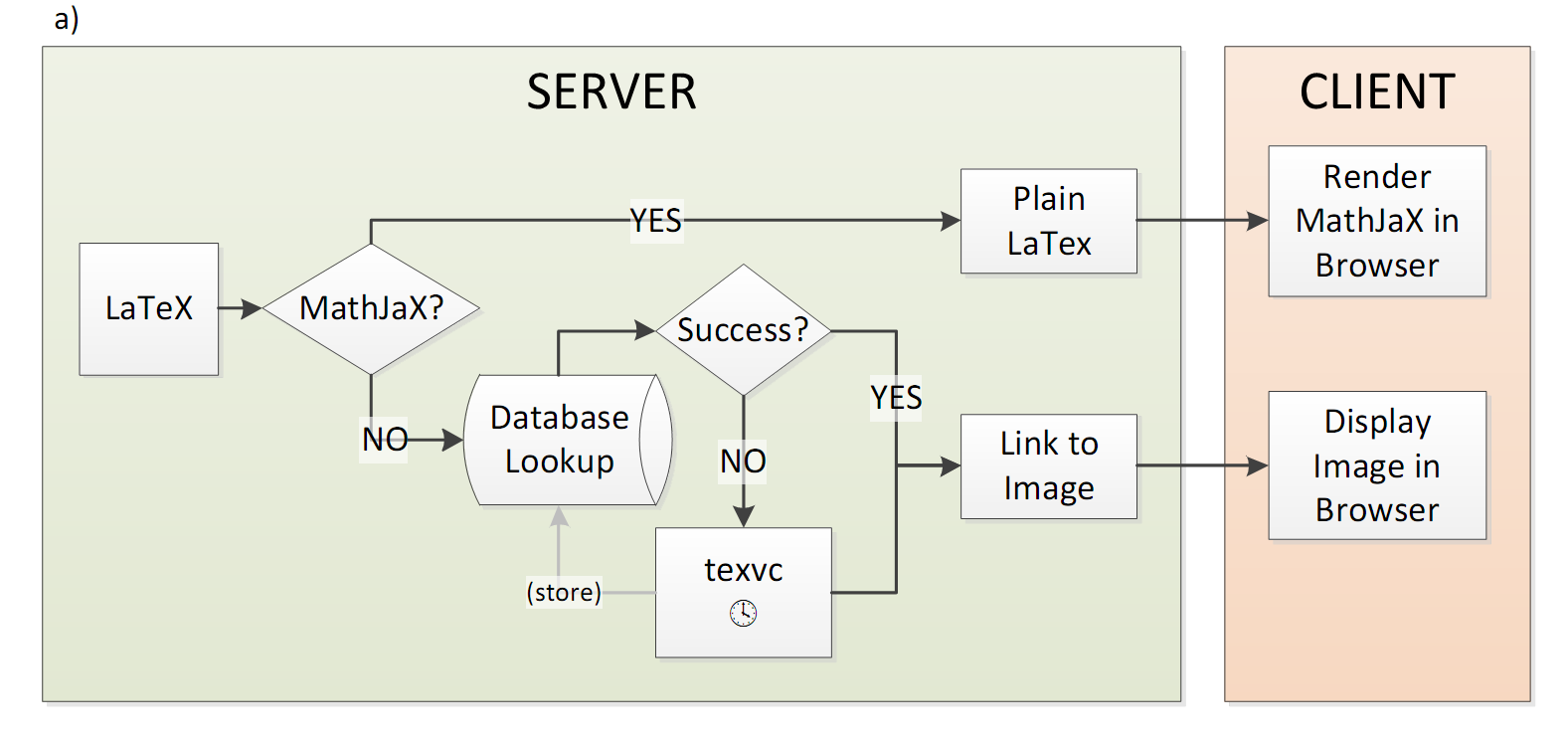}
\includegraphics[width=\textwidth]{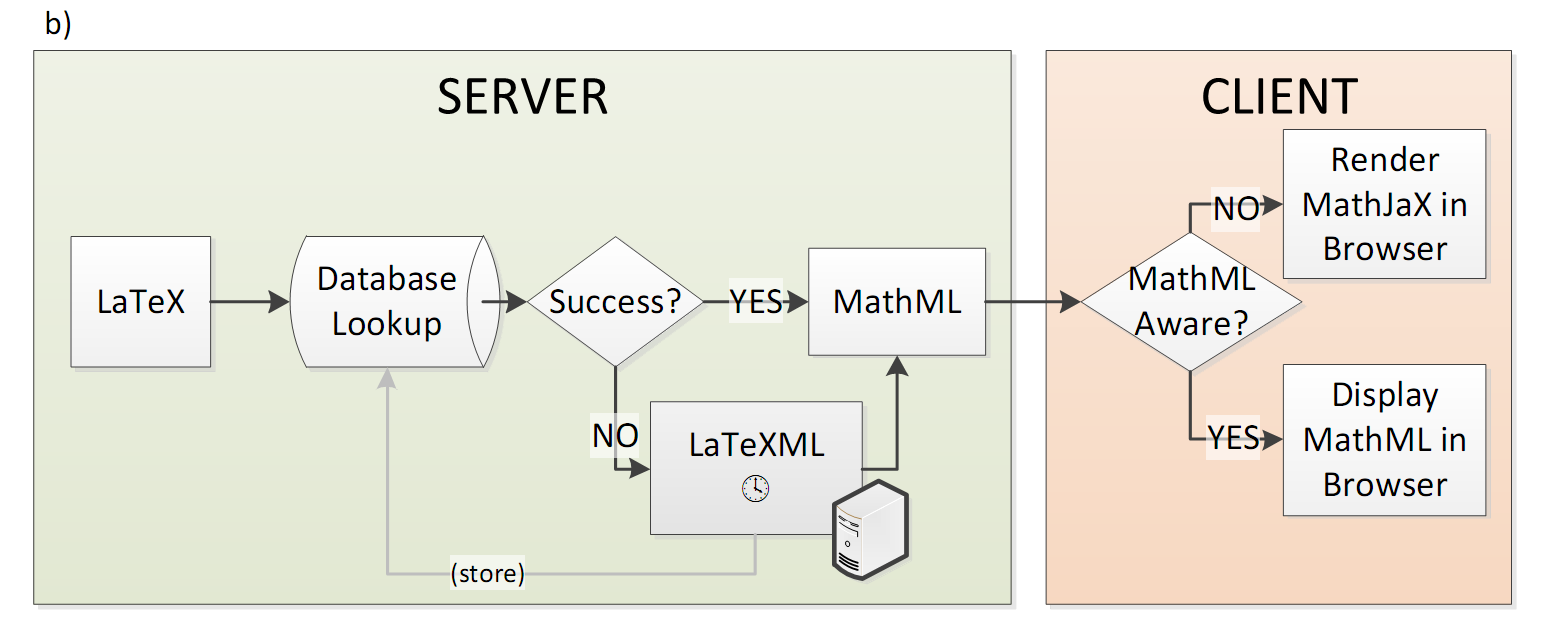}
	\caption{An overview of the \LaTeX{}   rendering process in the current version of MediaWiki (top) and the proposed new method (bottom).
	The clock symbol denotes that the rendering process takes about a second per formula on a typical desktop computer.
	The computer icon denotes that the \LaTeXML~rendering process of may  be processed on another server.}
	\label{fig.system}
\end{figure}

Today, the Wikimedia Foundation, as well as the majority of content providers that provide mathematical formulae, use images to display their mathematical content.
On the one hand this is a very robust solution, but on the other hand it is has serious disadvantages~\cite{Marchiori}.
Information retrieval is challenging using images alone.
To overcome these shortcomings \MathML{} (the Mathematical Markup Language) was developed to create mathematical notations suitable for websites~\cite{Miner2005,Youngen1997}.
However, neither content providers, nor web browser developers incorporated this standard in an adequate manner.
In fact, there are many browsers in use today that do not support \MathML.
As a consequence, the American Mathematical Society, the American Institute of Physics, Elsevier, IEEE and many other important players in this community support the MathJax project~\cite{Cervone2012} initiated in 2009.
The MathJax JavaScript library converts both the plain \LaTeX-source and \MathML{} to a format that may be displayed on almost any device and browser.
This is a major advantage, since it enables content providers to refrain from using the deprecated images.

From an information retrieval and math search point-of-view \MathML{} is better than \LaTeX-source, since it can be processed without compilation.
On the other hand \MathML{} was not designed to be written by humans.
Thus, a content provider who decides to use \MathML{} will need to convert the standard \LaTeX-input to \MathML.
In particular, for mathematical search it is highly recommended to use a converter that produces meaningful Content-\MathML\footnote{Content-\MathML{}  is a semantic representation of the operation contained in the formula.} as well as correct Presentation-\MathML\footnote{Presentation-\MathML{}  is the visual representation of the formula that is displayed by the browser.}.

Today, Wikipedia uses the texvc~\cite{texvc} renderer to convert all the \LaTeX{} formulae to images\footnote{
There were some approaches that use texvc to produce \MathML{} as well, but due to the missing capabilities in browsers to display \MathML{}, this attempt was not followed up a matter.}.
A current trend among the MediaWiki~\cite{MediaWiki} software development crowd 
is to bypass \LaTeX{}-rendering on the server side 
and to use MathJaX to convert the \LaTeX-source instead (see figure \ref{fig.system}-a).
Unfortunately, this leads to a slow page loads since all the \LaTeX{}   formulae have to be compiled by the client side and information concerning mathematical search or information extraction is unavailable on the server side.

Thus, we have changed the way mathematical formulae are processed by MediaWiki as follows (see figure \ref{fig.system}-b).
The texvc renderer that was installed on the local MediaWiki server has been replaced by an http-interface to one or more\footnote{A basic load balancer has been integrated.} \LaTeXML-daemons~\cite{Ginev2011,Miller:latexml} that might be installed on remote servers.
These \LaTeXML-daemons convert the \LaTeX{}  formula to \MathML{} that is stored in the central database and delivered to the client.
On the client side MathJax might be used, if the client browser is not capable of displaying \MathML{} out of the box. The source-code of our implementation is publically available in the \LaTeXML{} branch of the MediaWiki source repository. Detailed instructions on the installation may be found at \url{http://www.formulaeearchengine.com}.

In contrast to the other \LaTeX{} to \MathML{}  converters  that provide correct presentation-\MathML, \LaTeXML~ provides content-\MathML.
This content information becomes valuable if one wants to search for formulae\cite{StaGinDav:maacl09}.

\section{Combining Math Expressions and Text}
\begin{figure}[ht]
\includegraphics[width=\textwidth]{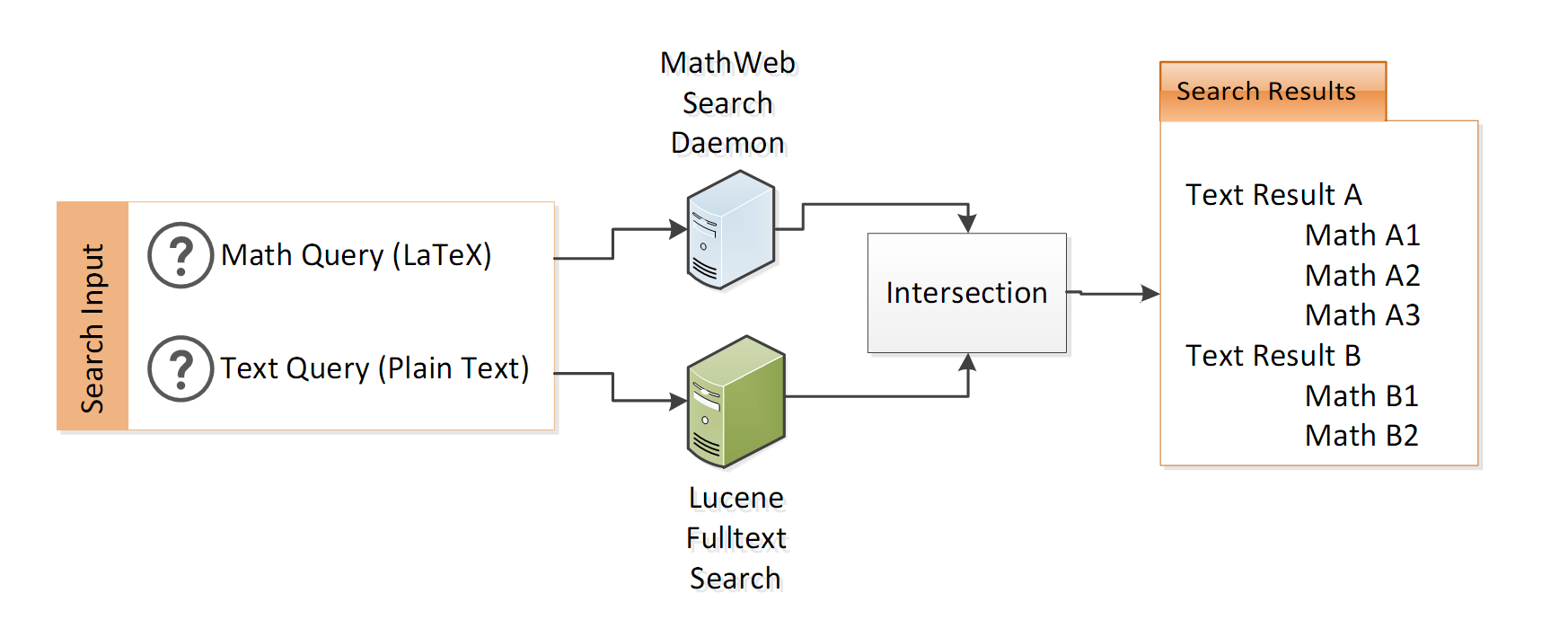}
	\caption{Illustration of the system architecture corresponding to the MathSearch plug-in for MediaWiki platform. Users specify a search query that consists of a text-query (e.g, Gr\"obner) and a math query (e.g $a?x^2+b?y^2+?z$) in a \LaTeX-fashion that might involve place-holder (e.g, $?x,?z,?z$). After the evaluation of the queries by the MathWebSearch Daemon and the Lucene Fulltext search respectively an intersection of the results is performed. All pages that match the text (An article that is about "Stable Normal Forms for Polynomial System Solving") are listed, conditioned on whether at least one formula matched the math query is result?(in this case, equation 16: $p_1=ax_1^2+bx_2^2+\epsilon_1x_1y_1$). The matching formula is then displayed as subitem of the text result preview. }
	\label{fig.search}
\end{figure}
In order to benefit from the knowledge conveyed in the formulae a specialized formula index is needed.
The logical first step would be to apply well known techniques from text search. This attempt was chosen for example by \cite{Misutka2011b,Misutka2008}. 
Due to the fact that there are many degrees of freedom in the notation of the formula (e.g., choice of the variables, order of terms, \dots) there are a couple of associated problems in practice. In Misutka and Galambos \cite{Misutka2008} treat these problems by applying seven unification rules.
However, this does not treat the root cause of the problem. 
 
More promising is the approach by Michael Kohlhase's group.
They have come up with a scalable~\cite{Kohlhase2012} search index for mathematical formula called MathWebSearch~\cite{KohSuc:asemf06}.
MathWebSearch indexes the content representation of the target formulae in a unified format, so that the degrees of freedom in the notation have been eliminated.
Although MathWebSearch meets these major user requirements in mathematical search, it leads to a large number of results that are not relevant to user queries. Therefore we have come up with a relatively naive but striking idea. We combine the MathWebSearch unification based mathematical search engine with a traditional full text search engine, that provides good precision and reduces the number of irrelevant results dramatically.

Making use of the \LaTeXML version of MediaWiki that contains formulae in content-\MathML{}-format, we build the open source MediaWiki site search expansion \textit{MathSearch}~\cite{MathSearch} that combines text and formula search (see Figure \ref{fig.search}) in the following way.
The existing MathWebSearch engine (that has no front-end) is combined with the text search engine Lucene~\cite{Mccandless2009}. We designed a simple front-end with two input fields. In one filed the user can specify a query for the mathematical content and the other optional field is reserved for the textual query. Both queries are then processed by the corresponding search engine then. After that the search results are grouped by text search result and the corresponding mathematical search results are classified as sub-items of the text results. A detailed description about the technical details and a demo may be found at \url{http://www.formulasearchengine.com}. 

\section{Experimental evaluation}
Our combined MathSearch solution was demonstrated at the MIR happening at CICM 2012. At this event our system competed with WebMIaS~\cite{Sojka2011}, that is based on canonicalized presentation \MathML{}, math trees and similarity search. The implementation of WebMIaS is based on Lucene. For math Indexing a specialized weighting function is used. At the MIR happening real human mathematicians post questions to the competitors in real time that have to submit the questions to their systems manually. 

As a data basis both systems have got 10'000 documents extracted from the ArXiv corpus two days in advance. Whereas WebMIaS hosted the documents on a remote server, we converted the documents to the Wiki mark-up and demonstrated our MathSearch solution on a virtual machine hosted on a standard laptop.

One of the tasks was to find
\begin{align}
B_{p+n} = B_n + B_{n+1} \bmod p \ \text{for all}\ n=0,1,2,\dots \nonumber
\end{align}
for the query $B_{p+n}$ both search engines came up with the correct result ranked at first position. WebMiaS found 455 additional results whereas our system just found one, since the search query didn't specify to use $\alpha$-conversions. 

For the combined text and math query \texttt{Gr\"obner}, $a?x^2+b?y^2+?z$ (see figure~\ref{fig.search}) our system came up with the expected solution. A detailed report from the MIR-happening by Professor J.H. Davenport will follow.

\section{Conclusion}
In this paper, we demonstrated how one can make use of the knowledge contained in the mathematical formulae by using a combined text and formula search interface. Furthermore, we showcased how content providers can publish their mathematical content in a modern and efficient way.
In the near future, we are going to apply advanced text mining methods to retrieve additional information about the symbols used in the formula and to resolve ambiguities.

\paragraph{Acknowledgments} Thanks to Deyan Ginev, Michael Kohlhase and Corneliu Prodescu for the fruitful discussions and Juan Soto for editing this paper.
   \bibliographystyle{amsplain}
   \bibliography{math_text_search}
\end{document}